
\magnification = 1200

\font\eightrm=cmr8
\font\eighti=cmmi8
\font\eightsy=cmsy8
\font\eightbf=cmbx8
\font\eighttt=cmtt8
\font\eightit=cmti8
\font\eightsl=cmsl8
\font\sixrm=cmr6
\font\sixi=cmmi6
\font\sixsy=cmsy6
\font\sixbf=cmbx6
\catcode`@11
\newskip\ttglue

\def\eightpoint{\def\rm{\fam0\eightrm}
\textfont0=\eightrm \scriptfont0=\sixrm \scriptscriptfont0=\fiverm
\textfont1=\eighti \scriptfont1=\sixi \scriptscriptfont1=\fivei
\textfont2=\eightsy \scriptfont2=\sixsy \scriptscriptfont2=\fivesy
\textfont3=\tenex \scriptfont3=\tenex \scriptscriptfont3=\tenex
\textfont\itfam=\eightit \def\it{\fam\itfam\eightit}
\textfont\slfam=\eightsl \def\sl{\fam\slfam\eightsl}
\textfont\ttfam=\eighttt \def\tt{\fam\ttfam\eighttt}
\textfont\bffam=\eightbf
\scriptfont\bffam=\sixbf
\scriptscriptfont\bffam=\fivebf \def\bf{\fam\bffam\eightbf}
\tt \ttglue=.5em plus.25em minus.15em
\normalbaselineskip=6pt
\setbox\strutbox=\hbox{\vrule height7pt width0pt depth2pt}
\let\sc=\sixrm \let\big=\eightbig \normalbaselines\rm}
\newinsert\footins
\def\newfoot#1{\let\@sf\empty
  \ifhmode\edef\@sf{\spacefactor\the\spacefactor}\fi
  #1\@sf\vfootnote{#1}}
\def\vfootnote#1{\insert\footins\bgroup\eightpoint
  \interlinepenalty\interfootnotelinepenalty
  \splittopskip\ht\strutbox 
  \splitmaxdepth\dp\strutbox \floatingpenalty\@MM
  \leftskip\z@skip \rightskip\z@skip
  \textindent{#1}\footstrut\futurelet\next\fo@t}
\def\fo@t{\ifcat\bgroup\noexpand\next \let\next\f@@t
  \else\let\next\f@t\fi \next}
\def\f@@t{\bgroup\aftergroup\@foot\let\next}
\def\f@t#1{#1\@foot}
\def\@foot{\strut\egroup}
\def\footstrut{\vbox to\splittopskip{}}
\skip\footins=\bigskipamount 
\count\footins=1000 
\dimen\footins=8in 

\def\ref#1{$^{#1}$}
\def\flex{\raise 6pt\hbox{$\leftrightarrow $}\! \! \! \! \! \! }

\newbox\bigstrutbox
\setbox\bigstrutbox=\hbox{\vrule height10pt depth5pt width0pt}
\def\bigstrut{\relax\ifmmode\copy\bigstrutbox\else\unhcopy\bigstrutbox\fi}
\def\refer[#1/#2]{ \item{#1} {{#2}} }
\def\rev<#1/#2/#3/#4>{{\it #1\/} {\bf#2}, {#3}({#4})}
\def\boxit#1{\vbox{\hrule\hbox{\vrule\kern3pt
\vbox{\kern3pt#1\kern3pt}\kern3pt\vrule}\hrule}}

\def\2figure#1#2#3#4{\vbox{ \hrule width#1truecm \hbox{\vrule height#2truecm
\hskip #1truecm
\vrule height#2truecm }\hrule width#1truecm \hbox{\vrule\vbox{\hsize #1truecm
\baselineskip=15pt
\noindent\strut#3}\vrule}\hrule width#1truecm
\hbox{\vrule\vbox{\hsize #1truecm
\baselineskip=15pt
\noindent\strut#4}\vrule}\hrule width#1truecm  }}
\def\3figure#1#2#3#4#5{\vbox{ \hrule width#1truecm \hbox{\vrule height#2truecm
\hskip #1truecm
\vrule height#2truecm }\hrule width#1truecm \hbox{\vrule\vbox{\hsize #1truecm
\baselineskip=15pt
\noindent\strut#3}\vrule}\hrule width#1truecm
 \hbox{\vrule\vbox{\hsize #1truecm
\baselineskip=15pt
\noindent\strut#4}\vrule}
\hrule width#1truecm \hbox{\vrule\vbox{\hsize #1truecm
\baselineskip=15pt
\noindent\strut#5}\vrule}\hrule width#1truecm  }}

\def\sqr#1#2{{\vcenter{\hrule height.#2pt
   \hbox{\vrule width.#2pt height#1pt \kern#1pt
    \vrule width.#2pt}
    \hrule height.#2pt}}}


\def\smin{\,\raise 0.06em \hbox{${\scriptstyle \in}$}\,}
\def\smsubset{\,\raise 0.06em \hbox{${\scriptstyle \subset}$}\,}

\def\Natural{\hbox{\hskip 1.5pt\hbox to 0pt{\hskip -2pt I\hss}N}}

\def\Rational{\hbox{\hbox to 0pt{\hskip 2.7pt \vrule height 6.5pt
                                  depth -0.2pt width 0.8pt \hss}Q}}
\def\Real{\hbox{\hskip 1.5pt\hbox to 0pt{\hskip -2pt I\hss}R}}
\def\Complex{\hbox{\hbox to 0pt{\hskip 2.7pt \vrule height 6.5pt
                                  depth -0.2pt width 0.8pt \hss}C}}

\magnification 1200

\def \sp {superparticle }

\def \cons {constraints }
\def \lb {\lbrace}
\def \rb {\rbrace}
\def \rbs {\rbrace^*}
\def \ps {p\!\!\!/}
\def \ns {n\!\!\!/}
\def \pts {\!\!\not\!\tilde p}
\def \nts {\!\!\not\!\tilde n}

\def \pits {{\tilde\Pi \!\!\!\!/}_{\sigma}}
\baselineskip .64cm
\centerline {{\bf A NEW CONSTRAINTS SEPARATION FOR THE ORIGINAL D=10}}
\vskip .65cm
\centerline {{\bf MASSLESS SUPERPARTICLE}}
\vskip .9cm
\centerline {D.Dalmazi\newfoot{*}{Partially
supported by CNPq}{}\newfoot{${}^{\dagger}$}{E-mail:
dalmazi@grt$0\!\!\!/0\!\!\!/0\!\!\!/$.uesp.ansp.br}}
\vskip .35cm
\centerline {UNESP - Campus de Guaratinguet\'a - DFQ -, CP 205 ,
Guaratinguet\'a - S.P. , Brazil.}
\vskip 2cm
\centerline {{\bf Abstract}}
\vskip .9cm
\noindent We study the problem of covariant
separation between first and second class constraints for the
$D=10$ Brink-Schwarz superparticle.
Opposite to the supersymmetric light-cone frame
separation, we show here that there is a Lorentz covariant way
to identify the second class constraints such that, however, supersymmetry
is broken. Consequences for the $D=10$ superstring are briefly discussed.
\vskip 5.5cm
\hfill {hep-th/9401011}
\vfill\eject

{\bf 1. - Introduction }
\vskip .5cm

There are basically two different formalisms for treating sypersymmetric
critical ($D=10$) strings, namely, the Neveau-Schwarz-Ramond (NSR)
formulation and the Green-Schwarz (GS) approach (see [1] for a review).
The GS superstring has the advantage of being explicitely spacetime
supersymmetric as opposite to the NSR string, also called spinning string,
where spacetime supersymmetry is only achieved after the GSO projection.

For the calculation of scattering amplitudes the manifest supersymmetry
of the GS approach represents, in principle, a great technical advantage
since some divergences which appear in the NSR formulation before the
GSO projection do not show up in the GS approach. In practice, however,
due to the lack of a Lorentz covariant quantization of the GS superstring
its advantages are withdrawned by the difficulties which appear
in the light-cone [2] (or semi-light-cone [3]) gauges where this
theory is usually quantized.

One of the difficulties of the noncovariant GS approach is that Lorentz
covariance of the amplitudes can only be proven indirectly by relating
the GS-superstring in the light-cone gauge with the light-cone
NSR string which is on its turn equivalent [4] to the covariant
NSR string. It is thus desirable to have an explicitly Lorentz covariant
quantization of the GS-superstring which, however, has become a formidable
task. In order to solve this problem we need a better undestanding of
a local fermionic symmetry of the GS-superstring which was discovered
by Siegel [5] and is called $\kappa$-invariance. Fortunately, this
symmetry also appears in the infinite tension limit of the GS
superstring, namely, the ten-dimensional massless superparticle [6].

In fact, all the problems which prevent a covariant quantization of the
GS superstring  are also present in the simpler case of the superparticle.
{}From the lagrangean point of view we have the following problems; on one hand
the algebra of gauge transformations only closes on shell and the
gauge generators are not linearly independent. On the other hand,
there is , apparently, no consistent Lorentz covariant gauge condition
to fix the $\kappa$-invariance. The first two problems do not
permit the use of the Faddeev-Popov procedure to fix the gauge, but
they can be circunvented [7] by applying the Batalin-Vilkowisky [8]
formalism according to which it is nescessary to introduce an infinite
set of ghosts of alternating statistics [7]. The problem of choosing
a consistent Lorentz covariant gauge condition is more dramatic and
it has not been solved yet.
In the hamiltonian formalism the problems appear when we try to identify
the second class constraints among the full set of constraints
(see, e.g., ref.[9]). There
seems to be no Lorentz covariant way of identifying the second
class constraints. Besides, the
first class constraints can only be  covariantly written in a
redundant (linearly dependent) way. In constructing the BRST charge
this redundancy leads to difficulties,
but for a Dirac quantization it does not represent a problem.

The apparently non-existence of both a Lorentz covariant gauge choice
and a Lorentz covariant identification of the second class constraints
has led many authors to suggest new formulations for the $D=10$
massless \sp and the $D=10$ superstring, see e.g. [9,10-18] and [19,20]
respectively. In [9,10-13] the basic idea was to extend the original superspace
by introducing pure gauge variables while, the most promising approach seems
to be the one initiated in [15] and further pursued in [18-20] where one tries
to convert part of the $\kappa$-invariance in worldline (worldsheet)
supersymmetry by the use of twistors and supertwistors variables, thus
making a direct connection between the NSR and the GS formulations.

In this paper we come back to the original superparticle [6]
and analyse it from the hamiltonian point of view.
Since we are only interested in a Lorentz covariant Dirac quantiation
we just have to face the constraints separation problem.
Without fixing any
gauge we show that besides the light-cone frame constraints
separation [9,21]
there is a Lorentz covariant separation which works as follows,
first of all we show that a light-like vector $n^{\mu}$
satisfying some conditions is all we need (see also [9]) for a consistent
Lorentz covariant identification of the second class constraints. Opposite
to [9] we do not introduce the vector $n^{\mu}$ by hand but construct it
(see formula (3.1))  out of the canonical pair ($x^{\mu},p^{\mu}$).
Although the second class \cons are identified in a redundant way
we are able to eliminate them by a kind of Dirac bracket. By using
such bracket we check that the Lorentz algebra closes classicaly although
supersymmetry and translation invariance are lost. We finally show
that for the  $D=10$ superstring the situation is different and
it is possible, in principle, to separate the constraints keeping
both supersymmetry and Lorentz invariance.
\vskip .5cm
{{\bf 2. - Global symmetries and constraints structure.}}
\vskip .5cm

The action for the massless superparticle [6] in ten dimensions can
be written in the first order formalism as:
$$
S= \int d\tau (p\cdot {\dot x} - {\it i}\theta\pts
{\dot\theta} -
{ep^2\over 2}) \quad , \eqno (2.1)$$
where ${\pts}_{\alpha\beta}=p_{\mu}({\tilde \sigma}^{\mu})_{\alpha\beta}$
($\alpha,\beta=1,2,\cdots,16 \, ; \,\mu =
0,1,\cdots,9$) and ${\theta}^{\alpha}$
is a right handed Majorana-Weyl spinor in ten dimensions, which has 16
real components. The real and symmetric matrices
$({\tilde \sigma}^{\mu})_{\alpha\beta}$ and $(\sigma^{\mu})^{\alpha\beta}$
satisfy
$$
\eqalignno{\left(\sigma^{\mu}{\tilde\sigma}^{\nu} +
\sigma^{\nu}\tilde\sigma^{\mu}\right)^{\alpha}_{\beta}&=\left(
\tilde\sigma^{\mu}{\sigma}^{\nu} +
\tilde\sigma^{\nu}\sigma^{\mu}\right)^{\alpha}_{\beta}=2\, \eta^{\mu\nu}
\delta^{\alpha}_{\beta} \; ,&(2.2a)\cr
Tr(\tilde\sigma^{\mu}{\sigma}^{\nu}) &= Tr(\sigma^{\mu}{\tilde\sigma}^{\nu})=
16\,\eta^{\mu\nu} \quad , &(2.2b)\cr
(\sigma^{\mu})^{(\alpha\beta}(\sigma_{\mu})^{\gamma\delta)}&=0=
(\tilde\sigma^{\mu})_{(\alpha\beta}(\tilde\sigma_{\mu})_{\gamma\delta)}
\quad , &(2.2c) \cr }
$$
where $\eta^{\mu\nu} = (-,+,\cdots,+)$ stands for the spacetime metric.

The global invariances of the action (2.1) consist of
superpoincar\'e transformations and dilatations which we call superweyl
transformations. Infinitesimally we have,
$$
\eqalign{
\delta x^{\mu} &= - 2\omega^{\mu\nu}x_{\nu} + a^{\mu} -
{\it i}\, \epsilon\,\tilde\sigma^{\mu}\theta + \lambda x^{\mu} \; ,\cr
\delta p^{\mu} &= - 2\omega^{\mu\nu}p_{\nu} - \lambda p^{\mu} \quad , \cr
\delta\theta^{\alpha}
&= {1\over 4}\omega^{\mu\nu}(\tilde\sigma_{\mu\nu}\theta)^{\alpha} +
\epsilon^{\alpha} + {\lambda \over 2}\theta^{\alpha} \; ,\cr
\delta e &= 2\lambda e \quad , }\eqno(2.3)
$$
where $\tilde\sigma_{\mu\nu} = \sigma_{\mu}\tilde\sigma_{\nu} -
\sigma_{\nu}\tilde\sigma_{\mu} \; (\sigma_{\mu\nu}=\tilde\sigma_{\mu}
\sigma_{\nu} - \tilde\sigma_{\nu}\sigma_{\mu}$).

Using the basic Poisson brackets\newfoot{$^{\sharp 1}$}{We use the
same symbol for Poisson brackets and Poisson anti-brackets
which are henceforth called generically Poisson brackets.}
$\lbrace x^{\mu},p_{\nu}\rbrace =
\delta^{\mu}_{\nu}\, ; \, \lbrace e, \Pi_e \rbrace = 1\, $ and
$\lbrace\theta^{\alpha},\Pi_{\beta}\rbrace = \delta^{\alpha}_{\beta},$
where $\Pi_{\beta} = {\partial^r {\cal L}\over \partial{\dot\theta}^{\beta}}$,
we can check that the superweyl transformations $(2.3)$ can be generated
by the following quantities:
$$
\eqalignno{J_{\mu\nu} &= x_{\nu}p_{\mu} - x_{\mu}p_{\nu} +
{1\over 4}\Pi\tilde\sigma_{\mu\nu}\theta\quad , &(2.4a)\cr
Q_{\alpha} &= \Pi_{\alpha}- {\it i} (\pts\theta)_{\alpha} \quad , &(2.4b)\cr
P_{\mu} &= p_{\mu} \qquad ,&(2.4c)\cr
D &= x\cdot p + {\Pi\theta\over 2} + 2\, e\,\Pi_e \quad , &(2.4d) \cr}
$$
which satisfy the superweyl algebra below:
$$
\eqalignno{\lb J_{\mu\nu},J_{\alpha\beta}\rb &=
\delta_{\mu\beta}J_{\nu\alpha} - \delta_{\mu\alpha}J_{\nu\beta} +
\delta_{\nu\alpha}J_{\mu\beta} - \delta_{\nu\beta}J_{\mu\alpha}
\; , &(2.5a)\cr
\lb J_{\mu\nu},P_{\alpha}\rb &= \delta_{\nu\alpha}P_{\mu} -
\delta_{\mu\alpha}P_{\nu} \quad , &(2.5b)\cr
\lb Q_{\alpha},Q_{\beta}\rb &= -2\,{\it i}\, (\pts)_{\alpha\beta}\quad ;
\quad \lb J_{\mu},Q_{\alpha}\rb =
- {(\sigma_{\mu\nu}Q)_{\alpha}\over 4}\; ,&(2.5c)\cr
\lb P_{\mu},D\rb &= -P_{\mu} \quad ; \quad\lb Q_{\alpha},D\rb = -
{1\over 2}Q_{\alpha}\; , &(2.5d) \cr }
$$
with the remaining brackets vanishing.

The superparticle
($2.1$) contains two bosonic first class
constraints :
$$\eqalignno{\varphi_1 \quad : \quad p^2 &\approx 0 \quad , &(2.6)\cr
\varphi_2 \quad : \quad\Pi_e &\approx 0 \quad , &(2.7) \cr}
$$
and the following 16 fermionic \cons .
$$
d_\alpha = \Pi_{\alpha} + {\it i}(\pts\theta)_{\alpha} = 0 \quad .\eqno (2.8)
$$
The only nonvanishing brackets among the \cons are the following ones
$$
\lb d_{\alpha},d_{\beta}\rb = 2{\it i} (\pts)_{\alpha\beta}\quad . \eqno(2.9)
$$
Due to (2.6) and the relation (2.2a) we get
$$
\lb d_{\alpha},(\ps d)^{\beta}\rb \approx 0 \quad . \eqno (2.10)
$$
Since $\lb p^2,(\ps d)^{\beta}\rb = 0 = \lb \Pi_e,(\ps d)^{\beta}\rb $ we
conclude that the combination $\ps d$, which is weakly equivalent to
$\ps\Pi$, corresponds to first class \cons . Indeed, the \cons
$\ps\Pi$ generate the local
$\kappa$-invariance [5] of (2.1), which reads
$$
\eqalign{\delta_{\kappa}\theta &= \ps \kappa (\tau) \quad ,\cr
\delta_{\kappa}x^{\mu} &=
-{\it i}\theta\pts\sigma^{\mu}\kappa(\tau) \quad ,\cr
\delta_{\kappa}e &= -2\,{\it i}\,\theta\dot\kappa (\tau) \quad , }\eqno (2.11)
$$
with $\kappa_{\alpha}$ a left-handed Majorana-Weyl spinor.

In order to understand the meaning of the remaining fermionic \cons
we suppose the existence of a vector $n^{\mu}$ satisfying
$$
p\cdot n \ne 0  \quad , \eqno(2.12)
$$
such that by use of (2.2a) we can write:
$$
d\; = \;{\cal P}_+\, d + {\cal P}_-\,d \quad , \eqno(2.13)
$$
where ${\cal P}_+ = (2n\cdot p)^{-1}\nts\ps\,$ and
$\,{\cal P}_- = (2n\cdot p)^{-1}\pts\ns\,$ .
By further assuming that $n^{\mu}$ is strongly light-like ($n^2=0$) we
show, using also (2.2b), that
$$
\eqalign{({\cal P}_{\pm})^2 &= {\cal P}_{\pm} \quad ; \quad {\cal P}_+{\cal
P}_-
    = 0 \quad , \cr
Tr \,({\cal P}_+) &= Tr \,({\cal P}_-) = 8 }\quad .\eqno(2.14)
$$
Therefore ${\cal P}_{\pm}$ are projection operators (strongly) and the 16
constraints $\;d_{\alpha}=0\;$ are equivalent to the 8 first class
constraints $\;\ps d=0\;$ and the 8 remaining constraints $\;\ns d=0\;$.
If the light-like vector $n^{\mu}$ satisfies
$$
\eqalignno{\lb n^{\mu},n^{\nu}\rb \; &= \; 0 \quad , &(2.15)\cr
\lb n^{\mu},d_{\alpha}\rb \; &= \; 0 \quad , &(2.16)\cr}
$$
then, we can proove that $\;\ns d\;$ represent 8 second class constraints
obeying:
$$
\lb(\ns d)^{\alpha},(\ns d)^{\beta}\rb \; = \; 4 \,{\it i}\, (n\cdot p)
(\ns)^{\alpha\beta}\quad . \eqno(2.17)
$$
So we come to the correct counting of independent constraints of the
$D=10$ massless \sp,
8 first class + 8 second class, which was deduced long ago [22]
in the light-cone frame where anyone of the light-cone directions
$\, n^{\mu}_{\pm} = (\mp 1,0,\cdots,0,1) \,$ satisfies (2.15),
(2.16) and $p^{\mu}$ can be
assumed to obey (2.12). In the next section we show that there is another
couple of light-like vectors $n^{\mu}_{\pm}$ which leads to
a Lorentz covariant separation still with the correct counting of
\cons .
\vfill\eject
{\bf 3. - Lorentz covariant constraints separation }
\vskip .5cm

We can always take appropriate nonlinear combinations of two
spacetime vectors to construct a couple of light-like vectors,
for instance, taking $x^{\mu}$ and $p^{\mu}$ we have
$$n_{\pm}^{\mu} \; = \; x^2 p^{\mu} - \lbrack x\cdot p \pm
((x\cdot p)^2 - x^2p^2)^{1\over 2}\rbrack x^{\mu} \quad , \eqno(3,1)$$
which identically satisfy $\; n^2_{\pm} = 0 \;$ . In order that
$n^{\mu}_{\pm}$ be real we have to impose the following scale
invariant restriction on the phase space:
$$
f\; \equiv \; (x\cdot p)^2 - x^2p^2 \; > \; 0 \quad . \eqno(3.2)
$$
The above restriction can only be violated by some noncausal configurations
inside the region where both $p^{\mu}$ and $x^{\mu}$ are space-like, i.e.,
$\; x^2>0 \;$ and $\; p^2>0 \;$. Notice also that (3.2) will be
automatically
satisfied on the physical states ($p^2\vert {\rm phys}>=0$).

Actually,
the restriction (3.2) is too strong, since we would have
real $n_{\pm}^{\mu}$ also for $\; f=0\;$, but if we intend
to use $n^{\mu}_+$ or $n^{\mu}_-$ to identify the second class constraints
we must have from (2.12) that $\; p\cdot n_{\pm} = -
f^{1\over 2}(f^{1\over 2}\pm x\cdot p) \ne 0 \;$ which
means that $\; f=0 \;$ must be discarded and we have to suppose either
$\; x\cdot p > 0 \,$ or $\, x\cdot p < 0\;$ if we choose
$n^{\mu}_+\,$ or $n^{\mu}_-\,$ respectively.
Henceforth we choose $n^{\mu}_+$ and assume the scale invariant restriction
$$
x\cdot p\; > \; 0 \quad , \eqno(3.3)
$$
which plays the role of the $\; p^+ > 0 \;$
condition of the light-cone frame.

The restrictions (3.2) and (3.3)
guarantee that the decomposition (2.13) holds and the 8
constraints\newfoot{$^{\sharp 2}$}{From now on we drop the subscript `+'
of $n^{\mu}_+\;$.} $\ns d$ are real such that we have
the correct counting of constraints. It is easy to check,
quite surprisingly,
that $n^{\mu}$ satisfies (2.15). The condition (2.16), however
is not obeyed and instead of (2.17) we get
$$
\eqalign{\lb (\ns d)^{\alpha}, (\ns d)^{\beta}\rb\; = \;
&4 {\it i} (\ns)^{\alpha\beta}\left( n\cdot p - {\Pi\theta\over 2}\right) +\cr
&+ \; {\rm terms}\;\,
{\rm proportional}\;\, {\rm to }\;\, \ns d \quad , \cr}\eqno(3.4)
$$
where we have used the identity
$$
(d\ns \tilde\sigma^{\mu}\theta) \sigma_{\mu}^{\alpha\beta} -
(\sigma_{\mu}d)^{(\alpha}(\sigma^{\mu} \nts \theta )^{\beta)}
\; = \; 2\, d\theta (\ns)^{\alpha\beta}\quad , \eqno(3.5)
$$
which is a direct consequence of (2.2c). Taking into account that
$\,\ns d\,$ has weakly vanishing Poisson brackets with the first
class constraints $\varphi_1\, ,\, \varphi_2\, ,\, \ps d$ and assuming
a third scale invariant restriction on the phase space :
$$
n\cdot p - {\Pi\theta\over 2}\; \ne \; 0\quad ,\eqno(3.6)
$$
which is also automatically satisfied on the physical states
(assuming the restrictions
(3.2) and (3.3)) where $\; \Pi\theta = d\theta = 0 \;$
we conclude that $\ns d$ really represent 8 independent second class
constraints which gives the correct counting of constraints once again.

Now we should notice that we cannot define the usual Dirac brackets
to eliminate $\ns d$ since the right-handed side of (3.4) has no
inverse due to the redundancy of the 16 linearly dependent constraints
$\ns d\;$. In such cases there is, {\it a priori }, no unique definition
of the bracket (see [23]) used
to eliminate the second class constraints, but by
requiring that no further restrictions on the phase space than (3.2),(3.3)
and (3.6) should be imposed we have just one possible definition, namely,
$$
\lb A,B \rbs \; = \; \lb A,B\rb -
\lb A,(\ns d)^{\alpha}\rb\,{(\pts)_{\alpha\beta}\over g}\,
\lb (\ns d)^{\beta},B\rb\quad ,\eqno(3.7)
$$
where $\; g=4\,{\it i}\,(n\cdot p)\,(2\,n\cdot p - \Pi\theta)\;$.
The Lorentz covariant definition (3.1) of $n^{\mu}$ assures the classical
closure of the Lorentz algebra for the constraints separation above. This
closure can be confirmed by an explicit computation of the
algebra of the superweyl generators $G_a$ (given in 2.4) in terms
of the bracket (3.7). By using the new brackets of the superspace
variables given in the appendix the reader can check that
$\;\lb G_a,G_b \rbs = \lb G_a,G_b\rb\;$ for most of the brackets (2.5)
except for the following ones
$$
\eqalignno{\lb P^{\mu},P^{\nu}\rbs \; &= \; {( x\cdot p +
f^{1\over 2}) ^2\over g}\, d\sigma_{\mu}\pts\sigma_{\nu}d\; ,&(3.8)\cr
\lb Q_{\alpha},Q_{\beta}\rbs \; &= \; \lb Q_{\alpha},Q_{\beta}\rb -
(\tilde\sigma_{\nu}\theta)_{\alpha}(\sigma_{\mu}\theta)_{\beta}\lb P^{\nu},
P^{\mu}\rbs\;&(3.9)\cr
\lb Q_{\alpha},P_{\mu}\rbs \; &= \; - (\tilde\sigma^{\nu}\theta)_{\alpha}
\lb P_{\nu},
P_{\mu}\rbs\quad &(3.10)\cr}
$$

Therefore, as in the ligh-cone frame separation, the superweyl algebra
will close only on the physical states ($\lb G_a,G_b \rbs
\vert{\rm phys}>=
\lb G_a,G_b \rb\vert{\rm phys}>$) where $d\vert{\rm phys}>=0\;$, but opposite
to the light-cone case we are able to keep Lorentz covariance
($\;\lb J_{\mu\nu},J_{\alpha\beta}\rbs = \lb J_{\mu\nu},J_{\alpha\beta}\rb\;$)
loosing translation invariance and supersymmetry. Our results lead to the
conjecture that we have to choose between Lorentz and supersymmetry invariance
for the massless superparticle since we have not found any other
self-consistent way to identify the second class constraints.

At this point it is important to comment upon the deep changes in the
canonical structure of the phase space (see appendix) caused by the
elimination of the constraints $\ns d$ with $n^{\mu}$ given in (3.1). Of
particular interest is the non-commutative nature of the variables $x^{\mu}$
(see (A.1)). It is remarkable that even on the physical states ($d=0$)
the coordinates $x^{\mu}$ have nonvanishing brackets among themselves, which
will lead, at the quantum level, to noncommuting position operators. Such
non-commutativity seems to be  a consequence (see remarks in [24])
of having only positive energy states ($E>0$) in the theory, which is
certainly true, in our case,
on the physical states where supersymmetry is unbroken.
It should be remarked that the
results found here and in the light-cone gauge [6] are in
agreement\newfoot{$^{\sharp 3}$}{Notice that although, our results are
classical as opposed to the quantum nature of the nonlocalization
problem the fact that $\lb x^{\mu},x^{\nu}\rbs \ne 0$ will clearly
survive the quantization.} with the
non-existence of Lorentz covariantly defined commuting position
operators for positive energy particles with nonzero spin ( nonlocalization
problem ) which was stablished long ago [26,27,28].
\vskip .5cm
{\bf 4. - The superstring case and final remarks}
\vskip .5cm
Now we show how the ideas of the last sections can be generalized for
the  superstring
case.

The D=10 GS superstring contains also 16 fermionic
constraints\newfoot{$^{\sharp 4}$}{The subscript $\sigma$ represents
a coordinate of a
point of the string and is not to be confused with the matrices
$\sigma_{\alpha}$}

$$
\; D_{\alpha} \, = \, \Pi_{\alpha} + {\it i}\, (\pits\theta)_{\alpha}
- {\it i}\, (\pts\theta)_{\alpha}\, = \, 0\quad , \eqno(4.1)
$$
where
$$
\eqalign{\Pi_{\sigma}^{\mu}\;&=\;\partial_{\sigma}x^{\mu}
+ \partial_{\sigma}\theta\tilde\sigma^{\mu}\theta\quad ,\cr
\Pi_{\alpha}\;&=\;{\partial^l{\cal L}_{GS}
\over\partial\theta^{\alpha}(\tau,\sigma)}\; ; \;
p^{\mu}={\partial{\cal L}_{GS}\over\partial x^{\mu}(\tau,\sigma)}}\eqno(4.2)
$$
with ${\cal L}_{GS}$ standing for the Green-Schwarz superstring
lagrangean (see [1]).
The constraints (4.1) have equal time Poisson brackets analogous to (2.9):
$$
\lb D_{\alpha}(\sigma),D_{\beta}(\tilde\sigma)\rb =
2\,{\it i}\,\delta(\sigma - \tilde\sigma)(\ps_s)_{\alpha\beta}\quad .
\eqno(4.3)
$$
where $\; p^{\mu}_s = -p^{\mu} + 2\,\Pi_{\sigma}^{\mu} -
\partial_{\sigma}x^{\mu}\;$ .

As in the case of the superparticle, due to the fact that
$\;p_s^2\approx 0 \;$ , the constraints $D_{\alpha}$ split
into 8 first class and 8 second class. Thus, once again
we have a decomposition like
(2.13) where $\ps_s D$ represent 8 first class constraints while
$\ns D$ stand for the remaining 8 second class constraints with
$n^{\mu}$ a strongly light-like vector such that $\, p_s\cdot n\ne 0\,$.
Now comes an important differemce to the superparticle's case, i.e., we have
for the superstring other possibilities for building $n^{\mu}$
than the Lorentz non-covariant light-cone directions and the
non-supersymmetric choice (3.1). Indeed, instead of (3.1) we can define
for the superstring, for instance, the light-like vectors
$$
n^{\mu}_{\pm}\; = \; \Pi_{\sigma}^2\,p^{\mu}_s -
\left\lbrack\Pi_{\sigma}\cdot p_s\,\pm\,\left( (\Pi_{\sigma}\cdot p_s)^2
\, -\,p_s^2\Pi_{\sigma}^2\right)\right\rbrack\,\Pi_{\sigma}^{\mu}\; .
\eqno(4.4)
$$
Due to the supersymmetric properties of $\Pi_{\sigma}^{\mu}$ and $p_s$
($\lb\Pi_{\sigma}^{\mu},Q_{\alpha}\rb =0
=\lb p_s,Q_{\alpha}\rb$), contrary to (3.1) , the
definition (4.4) leads to a supersymmetric constraints separation. This
and other possibilities are now being investigated. Actually, it is
possible to show that the separation based on (4.4) corresponds
exactly to the proposal of [25], so we
partially unravel the quite misteryous suggestion
of that reference.

Finally, it should be emphasized that the considerations made in this work
are pure classical and for the separations (3.1) and (4.4) the
Dirac brackets are rather involved (see appendix) which means that a correct
treatment of the very likely normal ordering problems will be
of fundamental importance.
\vfill\eject
\centerline {{\bf Acknowledgements}}
\vskip .4cm
I am grateful to Prof. V.O. Rivelles for discussions and
a careful reading of the manuscript. I also thank a discussion with
Prof. H.P. Nicolai on the projection operators ${\cal P}_{\pm}$.
Conversations with H. Boschi-Filho and A. de
Souza-Dutra are also acknowledged.

Finally I thank the financial support of DAAD/CAPES (Sandwich Fellowship)
during my stay at the II. Institut f\"ur Theoretische Physik
(Universit\"at Hamburg) where this work was initiated.
\vskip .65cm
\centerline {{\bf Appendix}}
\vskip .38cm
New brackets for the phase space variables :
$$
\eqalignno{\lb x^{\mu},x^{\nu}\rbs \, &= \, -2\,
{p\cdot n \over g}\,(\theta\tilde\sigma^{\mu}\ns \tilde\sigma^{\nu}\theta)\,
+ \,{x^2\over g}(d\sigma^{\lbrack\mu}\pts\ns\tilde
\sigma^{\nu\rbrack}\theta)&\cr
&\phantom{= \,} + \,
{x^4\over g}\, d\sigma^{\mu}\pts \sigma^{\nu}d\, \quad .&(A.1)\cr
\lb p^{\mu},p^{\nu}\rbs\, &= \, {(x\cdot p + f^{1\over 2})^2\over g}
(d\sigma^{\mu}\pts \sigma^{\nu}d)\quad . &(A.2)\cr
\lb x^{\mu},p_{\nu}\rbs \, &= \, \delta_{\nu}^{\mu}\; + &\cr
&\phantom{=\,}{(x\cdot p + f^{1\over 2})\over g}\left( -{\it i}\,
d\sigma_{\nu}\pts\ns\tilde\sigma^{\mu}\theta \, + \,
x^2\, d\sigma^{\mu}\pts \sigma_{\nu}d\right) \, . &(A.3)\cr
\lb \theta^{\alpha},\theta^{\beta}\rbs \, &= \,
{-\, 2n\cdot p\over g}(\ns)^{\alpha\beta} \quad . &(A.4)\cr
\lb \Pi_{\alpha},\Pi_{\beta}\rbs \, &= \,
{\, 2n\cdot p\over g}(\pts\ns\pts)_{\alpha\beta}\quad . &(A.5)\cr
\lb \theta^{\alpha},\Pi_{\beta}\rbs \, &= \, \delta_{\beta}^{\alpha} -
{\, 2\,{\it i}\,(n\cdot p)\over g}(\ns\pts)^{\alpha}_{\beta}\quad . &(A.6)\cr
\lb \theta^{\alpha},x_{\mu}\rbs \, &= \, {2\,{\it i}\,(n\cdot p)\over g}
(\ns\tilde\sigma_{\mu}\theta )^{\alpha} \, + \,
{x^2\over g}(\ns\pts\sigma_{\mu}d)^{\alpha}\, . &(A.7)\cr
\lb \theta^{\alpha},p_{\mu}\rbs \, &= \, {(x\cdot p + f^{1\over 2})\over g}
\, (\ns\pts\sigma_{\mu}d)^{\alpha}\quad . &(A.8)\cr
\lb \Pi_{\alpha},x_{\mu}\rbs\, &= \, {{\it i}\,
x^2\over 2}(\pts\ns\pts\sigma_{\mu}d)_{\alpha}\, - \, {2\, n\cdot p\over g}
(\pts\ns\tilde\sigma_{\mu}\theta)_{\alpha}\, . &(A.9)\cr
\lb \Pi_{\alpha},p_{\mu}\rbs\, &= \,
{\it i}\,{(x\cdot p + f^{1\over 2})\over g}(\pts\ns\pts\sigma_{\mu}d)_{\alpha}
\quad . &(A.10)\cr}
$$
where $f=(x\cdot p)^2 -x^2p^2$ and $g=4{\it i}\, n\cdot p\, (2\, n\cdot p
- \Pi\theta)\,$.
\vskip .8cm
\centerline {{\bf References}}
\vskip .5cm
\refer[[1]/M. B. Green, J. H. Schwarz and E. Witten, ``Superstring Theory''
(Cambridge U.P., Cambridge 1987).]

\refer[[2]/M. B. Green and J. H. Schwarz , Nucl.Phys {\bf B243} (1984) 475.]

\refer[[3]/S. Carlip , Nucl. Phys. {\bf B284} (1987) 365.]

\refer[[4]/K. Aoki, E. D'Hoker and D. H. Phong, Nucl. Phys. {\bf B342} (1990)
149.]

\refer[[5]/W. Siegel Phys. Lett. {\bf B 128} (1983) 397 .]

\refer[[6]/L. Brink and J.H. Schwarz, Phys. Lett. {bf B 100} (1981) 310.]

\refer[[7]/R. E. Kallosh, Phys. Lett. {\bf B195} (1987) 369.]

\refer[/R. E. Kallosh and A. Morozov, Int. J. of Mod. Phys. {\bf A3}
(1988) 1943.]

\refer[[8]/I.A. Batalin and G.A. Vilkowisky, Phys.Rev. {\bf D28}
(1983) 2567 ; Nucl.Phys. {\bf B234} (1984) 106.]

\refer[[9]/L. Brink, M. Henneaux, C.Teitelboim, Nucl.Phys. {\bf B293} (1987)
505.]

\refer[[10]/H. Aratyn, Phys. Rev. {\bf D39} (1989) 503.]

\refer[[11]/M. B. Green and C. M. Hull, Mod. Phys. Lett. {\bf A5}
(1990)  1399.]

\refer[[12]/R. Kallosh, Phys. Lett. {\bf B251} (1990) 134 .]

\refer[[13]/A. Mikovic, M. Rocek, W. Siegel, P. van Nieuwenhuizen, J. Yamron
and A. E. van de Ven, Phys. Lett. {\bf B235} (1990) 106.]

\refer[/F. Essler, M. Hatsuda, E. Laenen, W. Siegel, J. P. Yamron,
T. Kimura and A. Mikovic, Nucl. Phys. {\bf B364} (1991) 67 .]

\refer[[14]/E. Nissimov and S. Pacheva and S. Solomon, Phys. Lett. {\bf B189}
(1987) 57.]

\refer[[15]/D. P. Sorokin, V. I. Tkach, D. V. Volkov and A. A. Zheltukhin,
Phys. Lett. {\bf B216} (1989) 302.]

\refer[[16]/N. Berkovits, Nucl. Phys. {\bf B350} 193(1991).]

\refer[[17]/Y. Eisenberg, Phys. Lett. {\bf B276} 325(1992).]

\refer[[18]/A. Galperin and E. Sokatchev , Phys. Rev. {\bf D46} 714(1992).]

\refer[[19]/N. Berkovits, Nucl. Phys. {\bf B379} 96(1992).]

\refer[[20]/F. Delduc, A. Galperin, P. Howe and E. Sokatchev,
``A twistor formulation of the heterotic D=10 superstring with
manifest (8,0) worldsheet supersymmetry'', Bonn-preprint
HE-92-19 (July-1992).]

\refer[[21]/J. M. Evans, Class. and Quantum Grav. {\bf 7} 699(1990).]

\refer[[22]/I. Bengtsson and M. Cederwall, G\" oteborg preprint 84-21 (1984)
unpublished.]

\refer[[23]/A. Dresse, J. Fisch, M. Henneaux and C. Schomblond,
Phys. Lett. {\bf B210} 141(1988).]

\refer[[24]/I. Bengtsson and M. Cederwall, Nucl. Phys. {\bf B302} (1988) 81.]

\refer[/I. Bengtsson, M. Cederwall and N. Linden, Phys. Lett {\bf B203}
(1988)96.]

\refer[[25]/J. M. Evans, Phys.Lett. {\bf B233} (1989) 307 .]

\refer[[26]/M. H. I. Pryce, Proc. R. Soc. {\bf A195} (1948) 62.]

\refer[[27]/T. D. Newton and E. Wigner, Rev. Mod. Phys. {\bf 21} (1949) 400.]

\refer[[28]/D. J. Almond, Ann. Inst. Poincar\` e {\bf A19} (1973) 105 .]

\end